\documentclass{ws-ijmpd}

\usepackage{color}

\begin{document}

\markboth{Morozova and Ahmedov} {Quantum Interference Effects in
Slowly Rotating NUT Space-time}

%
\catchline{}{}{}{}{}
%

\title{Quantum Interference Effects in
Slowly Rotating NUT Space-time}

\author{V.S. MOROZOVA}

\address{Institute of Nuclear Physics and
    Ulugh Beg Astronomical Institute\\ Astronomicheskaya 33,
    Tashkent 100052, Uzbekistan\\
    The Abdus Salam International Centre
 for Theoretical Physics, 34014 Trieste, Italy \\
    moroz-vs@yahoo.com}

\author{B.J. AHMEDOV}

\address{Institute of Nuclear Physics and
    Ulugh Beg Astronomical Institute\\ Astronomicheskaya 33,
    Tashkent 100052, Uzbekistan\\
    The Abdus Salam International Centre
 for Theoretical Physics, 34014 Trieste, Italy \\
    ahmedov@astrin.uzsci.net}

\maketitle

\begin{history}
\received{Day Month Year} \revised{Day Month Year}
\end{history}

\begin{abstract}
General relativistic quantum interference effects in the slowly
rotating NUT space-time as the Sagnac effect and the phase shift
effect of interfering particle in neutron interferometer are
considered. It was found that in the case of the Sagnac effect the
influence of NUT parameter is becoming important due to the fact
that the angular velocity of the locally non rotating observer
must be larger than one in the Kerr space-time. In the case of
neutron interferometry it is found that due to the presence of
NUT-parameter an additional term in the phase shift of interfering
particle emerges. This term can be, in principle, detected by
sensitive interferometer and derived results can be further used
in experiments to detect the gravitomagnetic charge. Finally, as
an example, we apply the obtained results to the calculation of
the UCN (ultra-cold neutrons) energy level modification in the
slowly rotating NUT space-time.
\end{abstract}

\newpage

\section{Introduction}

    Due to the weakness of gravitational field in the Solar system,
many effects of general relativity predicted by theory are tiny
and not detected yet. Among them theoretical prediction about
existence of gravitomagnetic monopolar charge which was
theoretically discovered by Newman, Tamburino and
Unti~\cite{nut63} and called as NUT parameter.

According to Gravitoelectromagnetism~\cite{Ciu} there is analogy
between gravitation and electromagnetism in weak field and slow
motion approximation.  As electric currents in classical
electrodynamics create magnetic field, in the same way mass-energy
currents must produce gravitomagnetic fields. Verification of the
gravitomagnetism would test and prove the ideas of Einstein and
others that the local inertial frames are determined by the
distribution and currents of mass-energy throughout the Universe.
For some relevant information about research in this direction one
can look to~\cite{Ciu}-\cite{Manko}.

Moreover similar to magnetic Dirac's monopole there is hypothesis
about existence of gravitomagnetic monopole. Here we use slowly
rotating NUT metric for the exterior space-time of source with
nonvanishing NUT-parameter, related with gravitomagnetic charge,
as we did in our preceding papers~\cite{kaah,abahka}. In this
metric space-time we investigate two quantum interference
phenomena: so called Sagnac effect and the phenomena of phase
shift in neutron interferometer. It is connected with the fact
that quantum interference measurements based on macroscopic
quantum effects can provide sensitive precise accuracy to detect
weak gravitational effects. Our main goal here is to find
influence, caused by the existence of the NUT-parameter, on these
effects and prepare proposals for further experiments, which could
detect this influence.

The Sagnac effect is well known and thoroughly studied (see, for
example~\cite{Sag1}). It presents the fact that between light or
matter beam counter-propagating along a closed path in a rotating
interferometer a fringe shift $\Delta\varphi$ arises. This phase
shift can be interpreted as a time difference between the
propagation times of the co-rotating and counter-rotating beams.
The expression for time delay $\Delta T$ between two beams, as it
can be seen below, doesn't include the mass or energy of
particles. That is why we may consider the Sagnac effect as the
"universal" effect of the geometry of space-time, independent of
the physical nature of the interfering beams. Here we extend the
recent results obtained in the papers~\cite{Sag2,Sag3}, where it
has been shown a way of calculation of this effect in analogy with
the Aharonov-Bohm effect, to the case of slowly rotating NUT
space-time.

In addition to this technique we use here one more alternate way
to calculate the Sagnac effect, which is based on the idea
suggested in works~\cite{arifov70,arifov73} that the speed of
light depends on the direction of a light ray in space around
axial-symmetric gravitating object. Thus, we can explain the time
difference between two counter-propagating beams as a result of
the difference in speed in the different directions. We will use
this way to calculate the time delay and compare result with the
one already obtained by the previous method.

The experiment to test the effect of the gravitational field of
the Earth on the phase shift in a neutron interferometer were
first proposed by Overhauser and Colella~\cite{OC}. Then this
experiment was successfully performed by Collela, Overhauser and
Werner~\cite{COW}. After that, there were found other effects,
related with the phase shift of interfering particles. Among them
the effect due to the rotation of the Earth (the Coriolis
force)~\cite{Page,WSC}, which is the quantum mechanical analog of
the Sagnac effect, and the Lense-Thirring effect~\cite{LTh84},
which is a general relativistic effect due to the dragging of the
reference frames. Particle interferometry in
Pleba\'{n}ski--Demia\'{n}ski generalized black hole space--time is
presented in the recent work~\cite{Claus1}. Interesting effects,
caused by including the effects of particle's spin, or terms
violating the Einstein Equivalence Principle can be found
in~\cite{Claus2}-\cite{Claus3}.

In the paper~\cite{kkf} a unified way of study of the effects of
phase shift in neutron interferometer due to the various phenomena
was proposed. Here we extend this formalism to the case of the
slowly rotating NUT space-time in order to derive such phase shift
due to the existence of NUT-parameter.

Recently Granit experiment~\cite{UCN1} verified the quantization
of the energy level of ultra-cold neutrons (UCN) in the Earth's
gravity field and new, more precise experiments are planned to be
performed. Experiments with UCN have high accuracy and that is the
reason to look for verification of the gravitomagnetism in such
experiments. As an example we investigate modification of UCN
energy levels caused by the existence of NUT parameter.

Throughout we use a space-like signature $(-,+,+,+)$, Greek
indices are taken to run from 0 to 3, Latin indices run from 1 to
3.

The Kerr-NUT spacetime metric for the gravitomagnetic monopole in
the approximation of slowly rotating and weak gravitating source
looks as (see~\cite{exact82})
\begin{equation}
\label{metric} ds^2=-N^2dt^2+N^{-2}dr^2+r^2(d\theta^2+\sin^2\theta
d\phi^2)-2[\omega(r) r^2 \sin^2 \theta+2 l N^2 \cos\theta]\ dt \
d\phi\ ,
\end{equation}
where $N^2=(1-2M/r)$ is the lapse function, $M$ is the total mass
of the gravitating object, $\omega=2Ma/r^3$ is the angular
velocity of dragging of inertial frames, $a=J/M$ is the specific
angular momentum being equal to the total momentum $J$ of the
gravitating object per unit mass, $l$ is the NUT-parameter (we
assume that $c=G=1$, where $G$ is the Newtonian gravitational
constant, $c$ is the speed of light).

The paper is organized as follows. In section 2 we calculate the
Sagnac effect in the slowly rotating NUT space-time. In section 3
we find the expression for the phase shift in neutron
interferometer and UCN energy levels change caused by the
existence of NUT parameter. We also investigate the derived
expressions and compare them with already known terms. Section 4
is devoted to conclusions.

\section{The Sagnac effect}

As it was presented in the paper~\cite{Sag3} the Sagnac effect can
be calculated in analogy with the Aharonov-Bohm effect. In this
case the phase shift between two light or matter beams
counter-propagating in flat space-time along a closed path in a
rotating interferometer will be
\begin{equation}
\label{deltaphi}
 \Delta\varphi=\frac{2m u_0}{c\hbar}\oint_C \textbf{A}_G\cdot
d\textbf{x}\ .
\end{equation}

The corresponding time difference between the propagation times of
the co-rotating and counter-rotating beams will be equal to
\begin{equation}
\label{deltat} \Delta T=\frac{2 u_0}{c^3}\oint_C \textbf{A}_G\cdot
d\textbf{x}\ .
\end{equation}

In the formulae (\ref{deltaphi}) and (\ref{deltat}) $m$ is the
mass (or the energy) of the particle of the interfering beams,
$\textbf{A}_G$ is the gravito-magnetic vector potential, which is
obtained from the expression
\begin{equation}
\label{potential}
\textbf{A}^G_i\equiv c^2\frac{u_i}{u_0}\ ,
\end{equation}
and $u(x)$ is the unit four-velocity of particles:
\begin{equation}
u^\alpha\equiv\left\{\frac{1}{\sqrt{-g_{00}}},0,0,0\right\}\
,\quad u_\alpha\equiv\left\{-\sqrt{-g_{00}},g_{i0}u^0\right\}\ .
\end{equation}

Phase shift in the equation (\ref{deltaphi}) is detected by a
uniformly rotating interferometer and the time difference in
(\ref{deltat}) is measured by comoving observers.

Now we apply these equations to make the relevant calculations in
spacetime metric (\ref{metric}). It should be noticed that in the
equatorial plane $(\theta=\pi/2)$ influence of the NUT-part of the
metric to the effect will be zero. To see how the Sagnac
time-delay and phase shift will be changed due to presence of
NUT-parameter, let us consider the plane, where $\theta=\pi/4$, so
$\sin^2\theta=1/2$ and $\cos\theta=1/\sqrt{2}$.

After the coordinate transformation $\phi\rightarrow\phi+\Omega
t$, where $\Omega$ is the angular velocity of the gravitating
object the line element will look as
\begin{eqnarray}
\label{metric1}
ds^2&=&-\left(N^2-\frac{r^2\Omega^2}{2}+\Omega\omega r^2+
2\sqrt{2}lN^2\Omega\right)dt^2+N^{-2}dr^2+\frac{r^2}{2}d\phi^2
\nonumber\\
& +&2 \left( \frac{r^2(\Omega-\omega)}{2}-\sqrt{2}lN^2 \right)
d\phi dt\ .
\end{eqnarray}

From this equation one can immediately see, that the unit vector
field $u(x)$ along the trajectories $r=R=const$ will be
\begin{eqnarray}
&&u_0=-(u^0)^{-1}\ , \nonumber\\
&&u_{\phi}=\left(\frac{R^2(\Omega-\omega)}{2}-\sqrt{2}lN^2\right)u^0\
,
\end{eqnarray}
where a notation
\begin{equation}
u^0=\left(N^2-\frac{R^2\Omega^2}{2}+\Omega\omega
R^2+2\sqrt{2}lN^2\Omega\right)^{-1/2}
\end{equation}
is introduced.

Now, inserting the components of $u(x)$ into the equation
(\ref{potential}) one obtain
\begin{equation}
\textbf{A}^G_{\phi}=-\left(\frac{R^2(\Omega-\omega)}{2}-\sqrt{2}lN^2\right)(u^0)^2\
.
\end{equation}

Integrating vector potential, as it is shown in equations
(\ref{deltaphi}) and (\ref{deltat}), one can get the following
expressions for $\Delta\varphi$ and $\Delta T$ (here we returned
to the physical units):
\begin{equation}
\label{ABDelta phi} \Delta\varphi=\frac{4\pi
m}{\hbar}\left(\frac{R^2(\Omega-\omega)}{2}-\sqrt{2}clN^2\right)(u^0)^{2}\
,
\end{equation}
\begin{equation}
\label{ABDelta T}\Delta
T=\frac{4\pi}{c^2}\left(\frac{R^2(\Omega-\omega)}{2}-\sqrt{2}clN^2\right)(u^0)^{2}\
.
\end{equation}

Following to the paper~\cite{Sag3} one can find a critical angular
velocity $\bar{\Omega}$
\begin{equation}
\bar{\Omega}=\omega+\frac{2\sqrt{2}clN^2}{R^2}\ ,
\end{equation}
which corresponds to zero time delay $\Delta T=0$. $\bar{\Omega}$
is the angular velocity of the zero angular momentum observers
(ZAMO). As we can see, the term with NUT-parameter presents a
positive addition to this velocity, in other words, parameter
$\bar{\Omega}$ in Kerr-NUT space-time becomes larger with compare
to that in Kerr one.

{Alternate approach for interpretation of Sagnac effect
 is given in the papers~\cite{arifov70},~\cite{arifov73} which is based on
 anisotropy of  the speed of light. According to Einstein the definition
 of simultaneity can be derived from the
postulate of the constancy of the speed of light. However the
constancy and anisotropy of the speed of light is not obvious,
when one consider light propagation in arbitrary four-dimensional
Riemann space-time in general relativity. Moreover assumption
about the anisotropy of the speed of light is not obvious even in
inertial frames, because all laboratory methods of measuring the
speed of light establish anisotropy only then light is moving
"there and back" rather than only "there" or "back". Measurement
of the speed of light in one single direction requires the
establishment of the way to compare time intervals in different
points of space before making a measurement i.e. the definition of
simultaneity without including any knowledge about the speed of
light. Therefore the definition of simultaneity is primary and
underlie any physical measurement, whereas the value of the speed
of light is secondary and derived from this fundamental
definition.}

{Let $P_1$ and $P_2$ be the events on the world line of the clock
being at rest in the given frame, corresponding to emission and
registration of the light signal reflected (event $P'$) from the
mirror placed at slight distance from the clock. Each of these
events have fixed value of proper time $\tau(P_1)$, $\tau(P_2)$
and $\tau(P')$.}

{According to \cite{arifov73} an event $P$ being simultaneous to
$P'$ is defined as the event, moment of proper time of which
satisfies to the condition
\begin{equation}
\label{sim1}
\tau(P)=\frac{1}{2}[\tau(P_1)+\tau(P_3)]+\frac{1}{2}[\tau(P_1)-\tau(P_3)]
\frac{a_n
dx^n}{\sqrt{\left(g_{ik}-\frac{g_{0i}g_{0k}}{g_{00}}\right) dx^i
dx^k}}\ ,
\end{equation}
while Einstein's definition of simultaneity is
\begin{equation}
\tau(P)=\frac{1}{2}[\tau(P_1)+\tau(P_3)]\ .
\end{equation}}

{In equation (\ref{sim1}) $a_n$ are the covariant components of
the metric vector, defined as
\begin{equation}
\label{mvector}
 a_n=\frac{g_{0n}}{\sqrt{-g_{00}}}\ , a=\sqrt{\frac{g^{ik}a_i a_k}{1-g^{mj}a_m a_j}}\ .
\end{equation}}

{Using equation (\ref{sim1}) one can obtain expression for
interval of proper time between two arbitrary, but close to each
other events as
\begin{equation}
d\tau=\sqrt{-g_{00}}\left(dx^0+\frac{g_{0i}}{g_{00}}dx^i\right) +
a_idx^i
\end{equation}
and write the metric of space-time in the form
\begin{equation}
\label{Ametric} ds^2=-e^{-2\phi}dt^2+2e^{-\phi}a_idx^idt+dl^2 \ ,
\end{equation}
where
\begin{equation}
dl^2=h_{ik}dx^idx^k\ ,
\end{equation}
\begin{equation}
h_{ik}=g_{ik}-\frac{g_{0i}g_{0k}}{g_{00}}-a_ia_k\ .
\end{equation}}
{From equation (\ref{Ametric}) one can derive the expression for
the absolute value of the speed of light as
\begin{equation}
\label{speed}
v_c=\sqrt{-g_{00}}\left(-a\cos\alpha+\sqrt{1+a^2\cos^2\alpha}\right)\
,
\end{equation}
where $\alpha$ is the angle between the light ray and the metric
vector. } Consequently, the speed of light has the smallest value
when the light ray propagates in the direction along the metric
vector, and the largest value in the opposite direction. The speed
of light is constant in all directions, lying on the conical
surface with the top in a given point and the axis of symmetry
along the metric vector. From the equation (\ref{speed}) one can
see that the product of the values of the speed of light in the
opposite direction is constant in a given point and the following
equation takes place
\begin{equation}
\label{vel}
\frac{1}{-g_{00}}\cdot v_c\bar{v}_c=1\ ,
\end{equation}
where $\bar{v}_c$ is the absolute value of the vector
$\left(-\vec{v}_c\right)$.

Using these discourses, one can conclude that the time difference
between the counter-propagating light rays on a closed path is
equal to the integral
\begin{equation}
\label{Delta t} \Delta
t=\oint\left(\frac{1}{v_c}-\frac{1}{\bar{v}_c}\right)dl\ ,
\end{equation}
where $dl$ is the element of the path.

In axial symmetric space-times (in particular in one of the slowly
rotating source with the non-zero NUT-parameter) metric vector is
perpendicular to the axis of symmetry. For the light rays,
counter-propagating in the plane $\theta=\pi/4$, one {should}
choose the angles $\alpha$ being equal to $0$ and $\pi$ and
\begin{equation}
\label{vel1}
v_c(\alpha=\pi, 0)=\sqrt{-g_{00}}\left(\pm a+
\sqrt{1+a^2}\right)\ .
\end{equation}

{Making algebraic transformations in the equation (\ref{Delta t})
with help of equations (\ref{vel}) and (\ref{vel1}) one can find
\begin{equation}
\label{nDelta t} \Delta
t=\oint\frac{\bar{v}_c-v_c}{v_c\bar{v}_c}dl=
\oint\frac{\left(\bar{v}_c-v_c\right)}{-g_{00}}dl= 2\oint u^0 a
dl\ .
\end{equation}}
{In the space-time metric (\ref{metric1}) the single nonvanishing
component of the metric vector (\ref{mvector}) is
\begin{eqnarray}
\label{comp a}
a_3&=&u^0\left(\frac{r^2(\Omega-\omega)}{2}-\sqrt{2}lN^2\right)\ ,
\end{eqnarray}}
{and its absolute value is
\begin{equation}
\label{ABSa}
a=\frac{\frac{\sqrt{2}u^0}{r}\left(\frac{r^2(\Omega-\omega)}{2}-
\sqrt{2}lN^2\right)}{\left(1-\frac{2(u^0)^2}{r^2}\left(\frac{r^2
(\Omega-\omega)}{2}-\sqrt{2}lN^2\right)^2\right)^{1/2}}\ ,
\end{equation}}
{which in the linear approximation in angular velocity of rotation
and NUT-parameter takes simple form
\begin{equation}
a=\frac{\sqrt{2}u^0}{r}\left(\frac{r^2(\Omega-\omega)}{2}-
\sqrt{2}lN^2\right)\ .
\end{equation}}

{Inserting it into (\ref{nDelta t}) we eventually obtain an
expression for the time delay between two counter-propagating rays
in the Sagnac effect (in physical units)
\begin{equation}
\label{ArDelta t} \Delta t=\frac{\sqrt{2}\cdot4\pi}{c^2}
\left(\frac{R^2(\Omega-\omega)}{2}- \sqrt{2}clN^2\right) (u^0)^2 \
.
\end{equation}}

{Now we would like to underline that the Sagnac effect here is
calculated in the plane $\theta=\pi/4$ instead of traditional one
$\theta=\pi/2$ in order to find effect related to NUT-parameter
which vanishes in the equatorial plane $\theta=\pi/2$. Classical
experiment on Sagnac effect has been performed in the equatorial
plane being orthogonal to axis of rotation where both approaches
given by equations (\ref{ABDelta T}) and (\ref{ArDelta t}) would
give common result
\begin{equation}
\Delta T=\frac{4\pi R^2\Omega}{c^2}(u^0)^2\ .
\end{equation}
Therefore the difference between equations (\ref{ABDelta T}) and
(\ref{ArDelta t}) is due to calculations made in the plane
$\theta=\pi/4$.}

{It should be noticed that in equations (\ref{ABDelta T}) and
(\ref{ArDelta t}) the term arising from the Lense-Thirring effect
has spatial dependence as $1/r$, while the term produced by
non-zero NUT-parameter has no spatial dependence and will not
decay with the distance from the surface of the gravitating
object.}

\section{Phase shift}

Following to the discourses suggested in the work \cite{kkf} we
start from the covariant Klein-Gordon equation
\begin{equation}
\nabla^\mu\nabla_\mu\Phi-(mc/\hbar)^2\Phi=0\ ,
\end{equation}
define the wave function $\Phi$ of interfering particles as
\begin{equation}
\Phi=\Psi exp\left(-i\frac{mc^2}{\hbar}t\right)
\end{equation}
and neglect the terms of order $O((v/c)^2)$.

Using slowly rotating NUT metric (\ref{metric}) and making the
coordinate transformation $\phi\rightarrow\phi-\Omega t$ we obtain
the following expression for the Schr\"{o}dinger equation
\begin{equation}
\label{Schr}
 i\hbar\frac{\partial \Psi}{\partial
t}=-\frac{\hbar^2}{2m}\left[\frac{1}{r^2}\frac{\partial}{\partial
r}\left(r^2\frac{\partial}{\partial
r}\right)-\frac{L^2}{r^2\hbar^2}\right]\Psi-\frac{M m}{r}\Psi
 -\Omega L_z\Psi +\frac{2 Ma}{r^3}L_z\Psi + \frac{l
\cos\theta}{r^2\sin^2\theta}L_z\Psi\ ,
\end{equation}
where $L^2$, $L_z$ are the orbital angular momentum operators:
\begin{equation}
L^2=-\hbar^2\left[\frac{1}{\sin\theta}\frac{\partial}{\partial\theta}
\left(\sin\theta\frac{\partial}{\partial\theta}
\right)+\frac{1}{\sin^2\theta}\frac{\partial^2}{\partial\phi^2}\right]\
,
\end{equation}
\begin{equation}
L_z=-i\hbar\frac{\partial}{\partial\phi}\ .
\end{equation}

From the equation (\ref{Schr}) one can see that the Hamiltonian of
the particle in interferometer can be represented as a sum
\begin{equation}
H=H_0+H_1+H_2+H_3+H_4\ ,
\end{equation}
where
\begin{equation}
H_0=-\frac{\hbar^2}{2m}\frac{1}{r^2}\frac{\partial}{\partial
r}\left(r^2\frac{\partial}{\partial}\right)+\frac{L^2}{2mr^2}\
,\quad H_1=-\frac{M m}{r}\ , \quad H_2=-\Omega L_z\ ,\quad
H_3=\frac{2M a}{r^3}L_z\ .
\end{equation}
$H_0$ is the Hamiltonian for a freely propagating particle, $H_1$
is the Newtonian gravitational potential energy, $H_2$ is
concerned to the rotation of the gravitating source, $H_3$ is
related to the Lense-Thirring effect (dragging of the inertial
frames). The phase shift terms due to $H_1, H_2$ and $H_3$ are
\begin{eqnarray}
&&\beta_{grav}=\frac{m^2gS\lambda}{2\pi\hbar^2}\sin\phi\
,\quad \beta_{rot}\simeq\frac{2m \vec{\Omega} \cdot\textbf{S}}{\hbar}\ ,\nonumber\\
&&\beta_{drag}\simeq\frac{2m}{\hbar
R^3}\textbf{J}\cdot\left[\textbf{S}-3\left(\frac{\textbf{R}}{R}\cdot\textbf{S}\right)
\frac{\textbf{R}}{R}\right]\ ,
\end{eqnarray}
correspondingly.

Here $S=d_1d_2$ is the area of interferometer, \textbf{S} is the
area vector of the sector ABCD (see fig. 1),
$\vec{\Omega}=(0,0,\Omega)$ and \textbf{J}$=(0,0,J)$ are the
angular velocity and the angular momentum vectors of the object
correspondingly, $\textbf{R}$ is the position vector of the
instrument from the center of the gravitating object, $\lambda$ is
de Broglie wavelength.

The last term of the equation (\ref{Schr}) represents the part
$H_4$ of Hamiltonian
\begin{equation}
H_4=\frac{l \cos\theta}{r^2\sin^2\theta}L_z\
\end{equation}
related to the NUT-parameter.

Integrating it over time along the trajectory of the particle, one
can find the corresponding phase shift
\begin{equation}
\beta_4=\frac{1}{\hbar}\int\frac{l \cos\theta
}{r^2\sin^2\theta}L_zdt\ .
\end{equation}

Using a unit vector $\textbf{n}=(0,0,1)$, {presenting
$\textbf{r}=\textbf{R}+\textbf{r'}$, where $\textbf{r'}$ denotes
position of the given point of the interferometer from the center
of the instrument, and assuming that $\textbf{r'}/\textbf{R}$ is
small one can obtain that the angular dependence of $\beta_{NUT}$}
\begin{eqnarray}
\label{beta}  \beta_{NUT}&=& \beta_{4(ABD)}-\beta_{4(ACD)}=
-\frac{l m
 \cos\theta}{\hbar\sin^2\theta}\oint\frac{\textbf{n}\cdot(\textbf{r}\times
d\textbf{r})}{r^2}  \nonumber\\
&=&-\frac{l m \cos\theta}{\hbar
R^2\sin^2\theta}\textbf{n}\oint\frac{(\textbf{R}+\textbf{r'})\times
d\textbf{r'}}{|\textbf{R}+\textbf{r'}|^2} \simeq-\frac{l m
\cos\theta}{\hbar R^2\sin^2\theta}\left[\textbf{S}
-2\left(\frac{\textbf{R}}{R}\cdot\textbf{S}\right)\frac{\textbf{R}}{R}\right]
\textbf{n}\
\end{eqnarray}
has two singularities at $\theta=0$ and $\theta=\pi/2$. In further
calculations we assume $\theta=\pi/4$ and use for the expression
for angular dependence $\cos\theta/\sin^2{\theta}$ its value
$\sqrt{2}$.

Now we would compare this result with other terms, derived in the
paper~\cite{kkf} in order to evaluate its magnitude. As it was
found, $\beta_{drag}$ is the smallest term between others, which
has the similar structure as our $\beta_{NUT}$ and $10^{9}$ times
smaller then $\beta_{rot}$. But, in principle, it can be detected
with a sensitive "figure-eight" interferometer due to its
dependence on the distance from the center of the Earth as well as
the size and direction of interferometer (see~\cite{int}). Thus,
it is reasonable to consider the ratio $\beta_{NUT}/\beta_{drag}$.
If $\textbf{R}$ is perpendicular to $\textbf{A}$ then (here we
returned to physical units)
\begin{equation}
\label{ratio}
\frac{\beta_{NUT}}{\beta_{drag}}=\frac{5\sqrt{2}}{2}\frac{l c^3}{G
M R \Omega}\ .
\end{equation}

{For the Earth parameters $\Omega\sim 10^{-5} s^{-1}$, $M\approx
5.96\times 10^{27} g$ and $R\approx 6.4\times 10^8 cm$ one can
obtain
\begin{equation}
\frac{\beta_{NUT}}{\beta_{drag}}\sim 10^7 l \ ,
\end{equation}
which shows that detectable value of gravitomagnetic charge should
be at least $l\sim 10^6 cm$.}

\begin{figure}
\centerline{\psfig{figure=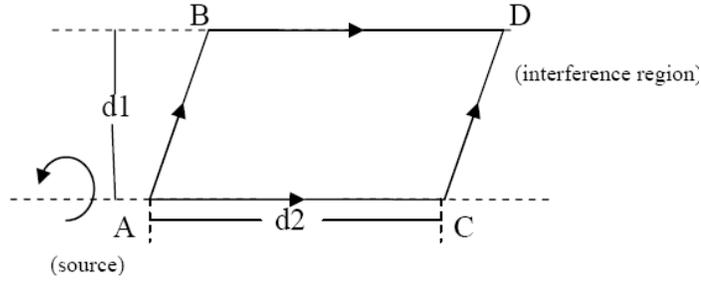,width=3.8in}}
\caption{Schematic illustration of alternate paths separated in
vertical direction in a neutron interferometer.} \label{fig1}
\end{figure}

Now we apply the obtained result for the Hamiltonian of the
particle, moving in a rotating space-time, endowed with the
gravitomagnetic charge, to the calculation of energy level of
ultra-cold neutrons (UCN) (as it was done for slowly rotating Kerr
space-time in the work~\cite{UCN}). In the paper~\cite{UCN} the
effect of the angular-momentum perturbation of the Hamiltonian
$H_2=-\Omega L_z$ on the energy levels of UCN is investigated. We
want to generalize this correction to the case of the gravitating
object (Earth in particular case) which possess also the NUT
parameter. Denote as $\psi$ the unperturbed non-relativistic
stationary state of the 2-spinor (describing UCN) in the field of
the rotating gravitating object with non-zero NUT-parameter. Then
we have
\begin{equation}
\label{H4psi}
H_4\psi=-i\hbar\frac{l\cos\theta}{r^2\sin^2\theta}\frac{\partial\psi}{\partial\phi}
= -i\hbar\frac{l\cos\theta}{r\sin\theta}\nabla\psi\cdot
\mathbf{e}_{\phi}\ ,
\end{equation}
where in the spherical coordinates Laplacian $\nabla\psi$ is
\begin{equation}
\nabla\psi = \frac{\partial\psi}{\partial r}\mathbf{e}_r +
\frac{1}{r}\frac{\partial\psi}{\partial\theta}\mathbf{e}_{\theta}
+
\frac{1}{r\sin\theta}\frac{\partial\psi}{\partial\phi}\mathbf{e}_{\phi}\
.
\end{equation}

Adopting new Cartesian coordinates $x,y,z$ with
$\mathbf{e}_x\equiv \mathbf{e}_{\phi}$ and axis $z$ being local
vertical, when the stationary state is assumed to have the form
\begin{equation}
\phi(\mathbf{x})=\phi_v(z)e^{i(k_1x+k_2y)} \ ,
\end{equation}
one can derive from (\ref{H4psi})
\begin{equation}
H_4\psi=-i\hbar\frac{l\cos\theta}{r\sin\theta}\frac{\partial\psi}{\partial
x} =\hbar k_1 \frac{l\cos\theta}{r\sin\theta}\psi = m u_1
\frac{l\cos\theta}{r\sin\theta}\psi \ ,
\end{equation}
with
\begin{equation}
u_1\equiv\mathbf{u}\cdot \mathbf{e}_\phi, \qquad
\mathbf{u}\equiv\hbar(k_1\mathbf{e}_x+k_2\mathbf{e}_y)/m\ .
\end{equation}

Following to the work~\cite{UCN} we can compute modification of
the energy level as a first-order perturbation:

\begin{equation}
\label{deltaE} (\delta E)_{NUT}\simeq(\psi|H_4\psi)= m
u_1\int\frac{l\cos\theta}{r\sin\theta}|\psi|^2 dV\ .
\end{equation}

Assume $r=(R+z)\cos\chi$ (where $\chi$ is the latitude angle) and
${\cos\theta}/{\sin\theta}$ to be equal to $1$, that is
$\theta=\pi/4$. Assuming now $z\ll R$ one can extend
(\ref{deltaE}) as
\begin{equation}
(\delta E)_{NUT}\simeq m u_1
\frac{l}{R\cos\chi}\int\left(1-\frac{z}{R}\right)|\psi|^2 dV = m u_1
\frac{l}{R\cos\chi}-m u_1 \frac{l}{R^2\cos\chi}\int z|\psi|^2 dV\ .
\end{equation}
Then we remember that $\int z|\psi|^2 dV$ is the average value
$<z>_n$ of $z$ for the stationary state $\psi=\psi_n$. For further
calculations we need to use formulae for $<z>_n$ from~\cite{UCN}

\begin{equation}
<z>_n=\frac{2}{3}\frac{E_n}{mg}\ .
\end{equation}

Now one can easily estimate the relative modification of the energy
level $E_n$ of the neutrons, placed in gravitomagnetic field, as
\begin{equation}
\frac{(\delta E)_{NUT}}{E_n}\simeq - \frac{2u_1l}{3g\cos\chi R^2}\ .
\end{equation}

We numerically estimate the obtained modification using the
following  parameters for the Earth: $u_1\simeq+10 m/s$,
$l\sim10^3cm$, $\cos\chi\simeq0.71$, $g\simeq10m/s^2$ and
$R\simeq6.4\times 10^8cm$. Then

\begin{equation}
\frac{(\delta E)_{NUT}}{E_n}\simeq 2.2\times 10^{-15} \ .
\end{equation}

From this one can see, that the influence of NUT parameter will be
stronger in the vicinity of compact gravitating objects with small
$R$.

\section{Conclusions}

In the present paper we have considered quantum interference
effects in slowly rotating Kerr-NUT space-time and found that the
presence of NUT-parameter in the metric can have influence on
different quantum phenomena. Namely, we obtained the phase shift
and time delay in Sagnac effect can be affected by monopolar
gravitomagnetic charge. Then, we found an expression for the phase
shift in a neutron interferometer due to existence of
NUT-parameter and concluded that it can be detected with the help
of  "figure-eight" interferometer. We also investigated the
application of obtained result to the calculation of energy levels
of UCN and found modifications to be rather small for the Earth,
but maybe more relevant for compact astrophysical objects.
Obtained information can be further used in experiments to detect
the interference effects related to the phenomena of
gravitomagnetism.

\section*{Acknowledgments}

VSM thanks the IUCAA for warm hospitality during her stay in
 Pune and AS-ICTP for the travel support. This research is
also supported in part by the UzFFR (projects 5-08 and 29-08) and
projects FA-F2-F079, FA-F2-F061 and A13-226 of the UzAS. This work
is partially supported by the ICTP through the OEA-PRJ-29 project
and the Regular Associateship grant. BJA acknowledges the partial
financial support from NATO through the reintegration grant
EAP.RIG.981259.

\end{document}